\documentstyle[twocolumn,aps]{revtex}

\begin{document}
\title{Polarization instabilities in a two-photon laser}
\author{Olivier Pfister,\cite{uva} William~J.~Brown,\cite{corvis}
Michael~D.~Stenner, and Daniel~J.~Gauthier\cite{ca}}
\address{Department of Physics and Center for Nonlinear and Complex Systems,\\
Duke University, Durham, North Carolina 27708-0305}
\date{March 6, 2001}
\wideabs{\maketitle

\begin{abstract}
We describe the operating characteristics of a new type of quantum
oscillator that is based on a two-photon stimulated emission process. This
two-photon laser consists of spin-polarized and laser-driven $^{39}$K atoms
placed in a high-finesse transverse-mode-degenerate optical resonator, and
produces a beam with a power of $\sim $0.2 $\mu $W at a wavelength of 770
nm. We observe complex dynamical instabilities of the state of polarization
of the two-photon laser, which are made possible by the atomic Zeeman
degeneracy. We conjecture that the laser could emit polarization-entangled
twin beams if this degeneracy is lifted.
\end{abstract}


}

The high degree of temporal coherence of laser light arises from a complex
interplay between the fundamental light-matter interactions of absorption,
spontaneous emission, and stimulated emission. The coherence properties of
the generated light, characterized by the temporal coherence function, can
be altered significantly and often in a surprising manner when quantum or
nonlinear optical effects are paramount. For example, the threshold,
stability, and quantum statistical properties of a single-atom maser \cite
{Walther} or laser \cite{Feld} operating in the `cavity quantum
electrodynamics regime' are very different from their multi-atom
counterparts. The coherence properties of laser light can also be modified
by exploiting different types of light-matter interactions. For example, the
two-photon laser \cite{EarlyTPL,Gauthier} is based on the higher-order
two-photon stimulated emission process, whereby two incident photons
stimulate an excited atom to a lower energy state and four photons are
scattered, as shown schematically in Fig.~\ref{setup}(a). This contrasts all
other `standard' lasers that are based on the one-photon stimulated emission
process whereby one photon stimulates an excited atom to a lower energy
state and two photons are scattered.

While replacing the standard one-photon stimulated emission process by a
high-order one might be expected to give rise to subtle differences
observable only at the quantum level \cite{Squeezing}, it has been predicted
that there will be dramatic changes in both the microscopic \cite{Quantum}
and macroscopic laser behavior even when many atoms participate in the
lasing process \cite{Dynamics}. The reason for these differences is that the
two-photon stimulated emission rate depends quadratically on the incident
photon flux, resulting in an inherently nonlinear light-matter interaction.

As an example, consider the effects of such a nonlinearity on the threshold
behavior of the two-photon laser. Briefly, the threshold condition for all
lasers is that the round-trip gain must equal the round-trip loss. For
one-photon lasers this yields the well known result that lasing will
commence when a uniquely defined minimum inversion density (proportional to
the gain) is attained via sufficient pumping. The situation is more
complicated for the two-photon laser because the unsaturated gain increases
with increasing inversion density {\em and} with increasing cavity photon
number. This results in a threshold condition specified by a uniquely
defined minimum inversion density and cavity photon number \cite{Concannon}
so that it cannot turn on unless quantum fluctuations \cite{Brune} or an
injected field \cite{Gauthier,Concannon} brings the intracavity light above
the critical value. In addition, once the minimum photon number exists in
the cavity, the photon number undergoes a run-away process, growing rapidly
until the two-photon transition is saturated \cite{Concannon}. Therefore,
the two-photon laser operates in the saturated regime (a source of optical
nonlinearity) even at the laser threshold, giving rise to the possibility
that the laser will display dynamical instabilities.

The primary purpose of this paper is to present experimental observations of
degenerate two-photon optical lasing that is based on laser-driven potassium
atoms \cite{tpg} contained in a very high finesse optical resonator. We
verify that the laser displays the expected turn-on behavior and observe
that it displays polarization instabilities. Such experiments provide a
window into the often conflicting predictions concerning two-photon lasers
and will help identify generic two-photon laser properties. In addition, we
suggest that this two-photon laser could emit polarization-entangled light
states of interest for quantum information applications.

Achieving two-photon lasing is hampered by the smallness of typical
two-photon stimulated emission rates and high saturation intensities, and
competing nonlinear optical processes often swamp or prevent the occurrence
of two-photon lasing. To overcome these difficulties, we follow the general
approach~\cite{Gauthier} of using laser-driven interactions to generate the
two-photon gain, and high-quality laser mirrors (developed for cavity
quantum electrodynamics experiments \cite{mirrors}) for the optical
resonator, as described below.

The two-photon gain medium we use in the experiments consists of a dense,
collimated, effusive beam of laser-driven $^{39}$K atoms that passes through
the center of the optical resonator orthogonal to the resonator axis, as
shown schematically in Fig.~\ref{setup}(b). We use the $D_{1}$ transition of 
$^{39}$K (6 MHz natural linewidth) whose Zeeman hyperfine structure is
displayed in Fig.~2. The atomic states are denoted by $|\alpha F_{\alpha
}M_{\alpha }\rangle $, with $\alpha =g$ for 4$^{2}S_{1/2}$ and $\alpha =e$
for 4$^{2}$P$_{1/2}$ levels. Quantum numbers $F$ and $M$ denote the total
angular momentum and its projection on the quantization axis $\hat{z}$,
defined by a weak magnetic field ($\leq 1$ Gauss) produced by a small
Helmholtz coil set around the cavity. The two-photon stimulated emission
rate is enhanced by the smallness of the ground state hyperfine splitting
for $^{39}$K of $\Delta _{g}/2\pi $=462 MHz. To create a population
inversion for multiphoton transitions starting from the state $|g22\rangle $%
, the atoms are continuously optically pumped into this state by two $\hat{%
\sigma}_{+}$-polarized fields whose frequencies are set close to $\omega
_{21}$ and $\omega _{22}$ (Bohr frequencies are denoted by $\omega
_{F_{e}F_{g}}$). We note that the homogenous linewidth of the Raman
two-photon transition is $\approx $6 MHz due to lifetime broadening of the
states by the optical pumping beams. The maximum atomic number density in
the resonator is $\sim 2\times 10^{11}$ atoms/cm$^{3}$, the atomic beam
diameter is 2.5 mm, and the measured residual Doppler absorption width is 30
MHz.

The two-photon laser resonator is placed in a vacuum chamber and consists of
two high-reflectivity ultra-low-loss mirrors of radius of curvature $R$=5 cm
set close together at a distance of $L$ = 1.464 cm in a linear
(standing-wave) transverse-mode-degenerate sub-confocal configuration. This
configuration suppresses normal one-photon lasing that can arise from
spectrally nearby gain features \cite{tpg}. The laser mirrors are mounted on
a super-invar structure to minimize thermal drift of $L$, with one mirror
attached to a mount actuated by three `picomotors' (New Focus MRA 8302) to
permit cavity alignment in vacuum and the other mirror is mounted on a
piezoelectric ceramic for fine scanning $L$. The mirrors have a
transmissivity of $T$ = 2$\times $10$^{-4}$ as measured by the manufacturer
(Research Electro-Optics). We find that the cavity finesse ${\cal F}%
=15,400\pm 150\simeq \pi /(T+A)$ using the ring-down technique, yielding a
loss per mirror $A=(4\pm 2)\times 10^{-6}$.

We use a short cavity length so that the free spectral range ($FSR=c/2L$
=10.24 GHz) is larger than the frequency difference between the two-photon
gain feature and all other nearby one-photon gain features, which are within
400 MHz of the two-photon frequency. We set the cavity length precisely to a
sub-confocal condition $L=R[1\pm \cos (\pi /p)]$ with $p=4$ so that the
transverse modes group into degenerate clusters, spaced by $c/2pL$=2.56 GHz
and centered on the longitudinal mode frequencies. Finally, to suppress
one-photon lasing on high-order transverse modes (which are no longer
frequency degenerate with the lower-order modes due to spherical aberrations 
\cite{Hercher}), we accurately position under vacuum 400 $\mu $m pinholes in
front of each cavity mirror and along the cavity axis using
picomotor-actuated translation stages. With the pinholes in place, ${\cal F}$%
=15,140, resulting in a longitudinal mode linewidth $FSR/{\cal F}=$ 680 kHz.
Within the cavity mode volume, we estimate that there are a maximum of 7$%
\times $10$^{6}$ atoms.

Two-photon stimulated emission into the laser cavity mode (frequency $\omega
_{l}$) occurs when the optically pumped atoms are illuminated by an intense $%
\hat{\sigma}_{-}$-polarized Raman pump beam of frequency $\omega _{d}$, and $%
\omega _{l}$ is adjusted to the two-photon transition frequency ($\simeq
\omega _{d}+\Delta _{g}/2$). Photons are added two at a time to the cavity
field via stimulated emission when two photons are annihilated from the $%
\hat{\sigma}_{-}$-polarized Raman pump field and the atom undergoes a
multi-photon transition from the state $|g22\rangle $ to any one of the
states $|g1M_{g}\rangle $, as shown in Fig. \ref{scatter}. Note that such a
hyper-Raman transition can be mapped onto the idealized two-photon
stimulated emission process shown in Fig. \ref{setup}(a) by working in a
basis in which the multi-level atomic structure is dressed by the Raman pump
field. The Raman pump beam has a maximum power of 300 mW and is focused to
an elliptical spot size (1/e field radius) of 3 mm along the cavity axis and
250 $\mu $m along the atomic beam axis, achieving a maximum intensity of
approximately 25 W/cm$^{2}$. Throughout the experiments, the pump beam
frequency is adjusted so that it is tuned to the blue side of the $\omega
_{22}$ transition by 512 MHz. The mutually orthogonal geometry of our
experiment suppresses competing phase-matched processes, ensuring that we
are constructing a true two-photon laser and not a parametric oscillator.

Our experimental system is very rich because there are 4 quantum pathways,
depicted in Fig.~\ref{scatter}, that connect the initial state $|g22\rangle $
and the final states $|g1M_{g}\rangle $, differ only by the state of
polarization of the cavity field, and are frequency degenerate at zero
magnetic field. The two-photon resonator can support any state of
polarization (it is measured to be highly isotropic) hence all pathways of
Fig.~\ref{scatter} must be considered.

Experimental evidence for the initiation of two-photon lasing is presented
in Fig.~\ref{trigger}, obtained using the following procedure. The atomic,
optical pumping, and Raman pump beam are turned on and a weak tunable
continuous-wave $\hat{z}$-polarized laser beam, produced by an auxiliary
laser, is injected into the two-photon laser resonator. The frequency of the
injected beam is scanned repeatedly through the cavity resonance while
monitoring the power emanating from the cavity, and the cavity frequency is
manually adjusted around the vicinity of the expected two-photon transition
frequency ($\omega _{l}\simeq \omega _{d}+\Delta _{g}/2$). We observe that
the emitted power increases when the cavity resonance is set to the
two-photon transition frequency and, for sufficiently high injected photon
number, two-photon lasing is initiated as indicated by the emitted power
remaining high even after the frequency of the injected light has swept
through the cavity resonance.

To perform more detailed measurements of the initiation process, we block
the auxiliary laser beam, set the cavity frequency to the peak of the
observed two-photon transition frequency, momentarily block the Raman pump
beam to quell any pre-existing two-photon lasing, and measure the power of
the light emitted from the two-photon laser resonator using a avalanche
photodiode (Hamamatsu C5460, 20 MHz bandwidth). In this situation, there is
essentially no light emitted from the resonator, indicating that quantum
fluctuations do not provide a sufficient number of photons in the cavity to
satisfy both lasing criteria. In an attempt to force the laser to the `on'
state, we typically inject a 1.2 $\mu $s-long pulse of light into the
resonator, which is gated from the auxiliary laser beam using an
acousto-optic modulator. Figure~\ref{trigger}(a) shows the case when a
below-threshold pulse is injected into the resonator. The dashed vertical
lines indicate the duration of the injected pulse and we estimate that $%
n_{inj}$=1.1$\times $10$^{5}$ photons are injected into the cavity. For
slightly higher trigger pulse powers [Fig.~\ref{trigger}(b), $n_{inj}$=2.2$%
\times $10$^{5}$], the cavity photon number grows substantially and remains
high for a few microseconds, but is still insufficient to initiate lasing.
The fact that the photon number stays high for a duration much longer than
the pulse width indicates that we are very close to satisfying both
two-photon laser threshold criteria. For $n_{inj}=$3.3$\times $10$^{5}$
[Fig.~\ref{trigger}(c)], the power emitted by the cavity rises to
approximately 0.2 $\mu $W and remains at this value, corresponding to an
average cavity photon number of 2.2$\times $10$^{6}$ and an intracavity
intensity of 7.3 W/cm$^{2}$ at the cavity waist. We observe that the power
emitted from the cavity remains high for up to a few seconds until the
frequency of the Raman pump or the cavity drifts off the two-photon
resonance or the Raman pump beam is blocked momentarily. Figure~\ref{trigger}%
(d) shows the power emitted from the resonator on a longer time scale for a
larger injected number of photons and during a different experimental run
but under essentially identical conditions. The experimental data of Fig.~%
\ref{trigger} gives convincing evidence that we have indeed observed
two-photon lasing.

As mentioned earlier, the degenerate hyperfine structure of $^{39}$K opens
up the possibility that the laser can operate on different states of
polarization. Recall that we trigger the laser to the `on' state using a $%
\hat{z}$-polarized beam. To gain some understanding of the subsequent state
of polarization, we place a linear polarizer oriented in the $\hat{z}$%
-direction in the beam and before the detector. We find that the two-photon
laser displays polarization instabilities even though the total emitted
power remains nearly constant on the time scale of the instability. It is
seen in Fig.~\ref{polarization}(a) that the state of polarization undergoes
very regular oscillations of period 0.11 $\mu $s, with a 50\% depth of
modulation. Similar dynamical behavior is observed for all polarizer
orientations, suggesting that the state of polarization is elliptical with
an ellipticity of 0.5 and a rotating major axis. Additional experiments
using multiple detectors to simultaneously record the power emitted in
different states of polarization are needed to accurately quantify the
laser's polarization state.

We find that the laser dynamics is quite sensitive to the applied magnetic
field: an increase by as little as 0.5 Gauss is sufficient to generate a
much more complicated pattern. Figure \ref{polarization}(b) shows the
complex, possibly chaotic behavior observed for a magnetic field strength of
2.0 G. \ An analysis of the data shown in Fig. \ref{polarization}(b) during
the interval after which the injected pulse is turned off reveals that the
irregular oscillations are characterized by a broad power spectrum and a
correlation function that drops to near zero within 100 ns; necessary, but
not sufficient conditions for the existence of chaotic oscillations. \
Determining whether the oscillations are chaotic will require a much longer
data set analyzed using nonlinear dynamics methods \cite{abarbanel}.

The observation of instabilities in the Raman two-photon laser is
surprising. Theoretical models of generic two-photon lasers, based on
simplified energy level schemes, predict that the laser should be stable in
the so-called `good cavity limit' \cite{Dynamics}. Our experiment is carried
out in this regime, where the cavity linewidth is narrower than, but
comparable to the population and atomic coherence linewidths. We suggest
three possible mechanisms that may be responsible for the observed
instability. One mechanism arises from the multiple frequency-degenerate
final lasing states. As the laser turns on and begins to saturate the
pathway shown in Fig.~\ref{scatter}(a) (because we inject a $\hat{z}$%
-polarized trigger field), an $\hat{x}$-polarized field can begin to grow on
the unsaturated pathways shown in Fig.~\ref{scatter}(b) and (c) (that is
enhanced by the existing $\hat{z}$-polarized cavity field). \ We note the
one-photon lasers with a continuously injected signal \cite{injected} or
with vector degrees-of-freedom can exhibit instabilities even in the `good
cavity limit' \cite{goodcavity}.

Another possible mechanism is that the lasing takes place on multiple
non-frequency-degenerate transverse modes arising from cavity mirror
birefringence, for example. \ We do not believe that this is likely because:
we observe that a linearly polarized laser beam passing through the cavity
in the absence of atoms remains linear to within our measurement sensitivity
(one part in 10$^{5}$) for various input states of polarization, the
transverse profile of the laser beam appears to be a lowest order TEM$_{00}$
mode as measured by a video camera, and the pinholes placed in the cavity
suppress all but the lowest-order transverse modes (we estimate the
diffraction loss of the TEM$_{01}$ mode of the empty cavity due to the
pinholes is about a factor of 10 greater than the loss of the TEM$_{00}$
mode \cite{Li}). \ Additional high-speed measurements of the
spatial-temporal behavior of the beam profile will be needed to completely
rule out the presence of transverse-mode instabilities.

Finally, the presence of standing waves may induce the instabilities; it is
well known that counterpropagating laser beams in conjunction with a tensor
nonlinear optical interaction can give rise to polarization instabilities
with a reduced instability threshold \cite{polarizationchaos}.
Distinguishing between the first and the third mechanisms and possibly
suppressing the instabilities should be possible using a stronger magnetic
field to lift the degeneracy of the different quantum pathways.

Lifting the degeneracy between the pathways also has the promise of yielding
polarization entanglement. Consider, for simplicity, a single emission
event. When the Zeeman levels are degenerate, as in Fig.~\ref{scatter}, the
superposition of all possible amplitudes gives a photon pair (1,2) in the
state $[\alpha _{1}|\,\hat{z}\,\rangle _{1}+\beta _{1}|\,\hat{x}\,\rangle
_{1}][\alpha _{2}|\,\hat{z}\,\rangle _{2}+\beta _{2}|\,\hat{x}\,\rangle _{2}]
$ with $|\alpha _{i}^{2}|$ +$|\beta _{i}^{2}|$=1, where $|\,\hat{\varepsilon}%
\,\rangle _{i}$ ($\varepsilon =x,y$) is the polarization-labeled Fock state
of the $i^{th}$ photon of the emitted pair. In a strong magnetic field, the $%
\hat{x}\hat{z}$ and $\hat{z}\hat{x}$ pairs become off-resonant and should
stop lasing, creating the polarization-entangled state $\alpha _{1}\alpha
_{2}|\,\hat{z}\,\rangle _{1}|\,\hat{z}\,\rangle _{2}+\beta _{1}\beta _{2}|\,%
\hat{x}\,\rangle _{1}|\,\hat{x}\,\rangle _{2}$. Maximal entanglement ($%
\alpha _{1}\alpha _{2}$= $\beta _{1}\beta _{2}$) can be achieved by varying
the resonant enhancement of a pathway, {\em e.g.}\ varying the Raman pump's
frequency and power, and the magnetic field.

In conclusion, we have described the design and characteristics of the first
Raman two-photon laser. The multilevel nature of the two-photon transition
allows interesting effects such as polarization instabilities and possible
photon entanglement. This opens up exciting lines of study in nonlinear
laser dynamics and quantum optics.

This work is supported by NSF grant PHY-9876988.

\begin{figure}[tbp]
\caption{(a) The idealized two-photon stimulated emission process. (b)
Two-photon laser experimental setup.}
\label{setup}
\end{figure}

\begin{figure}[tbp]
\caption{Two-photon Raman scattering diagrams showing the quantum pathways
for the possible states of polarization of the two-photon laser field (solid
line) and the cicularly polarized Raman pump beam (dashed line).}
\label{scatter}
\end{figure}

\begin{figure}[tbp]
\caption{Triggering the Raman two-photon laser by injecting a external
pulse. In (a) and (b), the injected field is not strong enough to drive the
laser to the high-power (`on') state.}
\label{trigger}
\end{figure}

\begin{figure}[tbp]
\caption{Polarization instabilities in the Raman two-photon laser for (a)
weak and (b) strong magnetic field.}
\label{polarization}
\end{figure}

\end{document}